\documentclass[prb,twocolumn]{revtex4-1}

\usepackage{amsmath,amsfonts,graphicx}

\newcommand{\half}{\tfrac{1}{2}}
\newcommand{\ev}[1]{\langle #1 \rangle}

\begin{document}

\title{Teaching renormalization, scaling, and universality \\
with an example from quantum mechanics}

\author{Steve T. Paik}
\email{{\tt paik_steve@smc.edu}}
\affiliation{Physical Science Department, Santa Monica College,
Santa Monica, CA 90405}

\date{\today}

\begin{abstract}
We discuss the quantum mechanics of a particle restricted to the half-line 
$x > 0$ with potential energy $V = \alpha/x^2$ for $-1/4 < \alpha < 0$.
It is known that two scale-invariant theories may be defined. By
regularizing the near-origin behavior of the potential by a finite square
well with variable width $b$ and depth $g$, it is shown how these two
scale-invariant theories occupy fixed points in the resulting $(b,g)$-space of
Hamiltonians. A renormalization group flow exists in this space and scaling 
variables are shown to exist in a neighborhood of the fixed points. 
Consequently, the propagator of the regulated theory enjoys homogeneous 
scaling laws close to the
fixed points. Using renormalization group arguments it is possible to
discern the functional form of the propagator for long distances and long
imaginary times, thus demonstrating the extent to which fixed points control
the behavior of the cut-off theory.

By keeping the width fixed and varying only the well depth, we show how the
mean position of a bound state diverges as $g$ approaches a critical value. 
It is proven that the exponent characterizing the divergence is universal in 
the sense that its value is independent of the choice of regulator.

Two classical interpretations of the results are discussed: standard Brownian
motion on the real line, and the free energy of a certain one-dimensional 
chain of particles with prescribed boundary conditions. In the former example,
$V$ appears as part of an expectation value in the Feynman--Kac formula. 
In the latter example, $V$
appears as the background potential for the chain, and the loss of extensivity 
is dictated by a universal power law.
\end{abstract}

\maketitle

\section{Introduction}

The inverse-square potential in quantum mechanics has a rich history that
continues to be updated in light of new connections to diverse physical 
phenomena including electron capture by neutral polar molecules,\cite{Camblong} 
the Efimov effect in a system of three identical bosons,\cite{Efimov} 
the transition between asymptotically free and conformal phases in QCD-like 
theories as a function of the ratio of the number of quark flavors to colors, 
and the AdS/CFT correspondence.\cite{Son}
A common thread running through these applications, and the reason for our
interest, is that the attractive $1/x^2$ potential is a 
fascinating case study that naturally calls upon 
the framework of the renormalization group (RG). We will see that the 
renormalization group approach mirrors, 
in many ways, the modern treatment of quantum effective field
theories whereby one demands that long-distance observables remain insensitive 
to the adjustment of fine details at short distances.

The purpose of this article is twofold. It is primarily a pedagogical treatment
of renormalization in the context of single-particle quantum mechanics 
and it is intended for teachers of quantum field theory. 
We feel that the example 
presented herein can provide an instructive introduction to basic RG ideas and
terminology. Because our renormalization group analysis involves a quantum
theory where one maintains full nonperturbative control, teachers may find 
that it serves as a useful aid for the beginning graduate student who is 
learning field theoretic
renormalization, but having difficulty separating the core principles from 
the technology required to do perturbative renormalization with many degrees of
freedom. We should mention that other pedagogical presentations of the 
renormalization of the inverse-square potential exist. Readers may wish to 
consult Refs.~\onlinecite{Essin} and \onlinecite{CoonHolstein} 
for additional background. However, in these works they study the case
$\alpha < -1/4$.

The other purpose of the article is to continue exploring the
implications of the renormalization group results uncovered in the work of
Kaplan, Lee, Son, and Stephanov in Ref.~\onlinecite{Son}. Although their work
is certainly not the first instance in which the inverse-square potential is
discussed, it is notable for, among other things, discussions of the beta 
function, operator anomalous dimensions at the fixed points, 
and the general phenomenon of conformal to non-conformal phase transitions.
In particular, we follow up their understanding of the
fixed-point structure with a natural extension to the quantum mechanical
propagator, and we use well-known quantum-classical equivalences to extract 
statistical lessons for specific one-dimensional classical systems.

Consider a particle in one spatial dimension subject to the potential
\begin{equation}
\label{V}
  V(x) = \Bigl\{\begin{array}{ll}
  \infty & x \leq 0 \\
  \alpha/x^2 & x > 0
  \end{array}  
\end{equation} 
for $-1/4 < \alpha < 0$.
We work in units where $\hbar = 1$ so that energy is the reciprocal of
time, and $\hbar^2/2m = 1$ so that energy is also the reciprocal of 
length-squared. This means that time and length-squared have equivalent
dimensions. In these units, $\alpha$ is a dimensionless number whose value
we do not imagine changing in any of our analyses.

The paper is organized as follows.
In Sec.~\ref{sec:pure} we explain why the choice $-1/4 < \alpha < 0$,
although giving perfectly consistent dynamics, is still peculiar. We then
explain that two distinct Hilbert spaces may be defined. In 
Sec.~\ref{sec:break} the Hamiltonian is modified at short-distances at the
expense of a dimensionful length scale and a coupling constant 
so that these two Hilbert spaces meld into one. In Sec.~\ref{sec:renormalize}
we discuss how the process of renormalization allows one to continuously
vary the short-distance modification without affecting a long-distance
observable. This naturally leads to the construction of the propagator in
Sec.~\ref{sec:prop} and analysis of its properties in Sec.~\ref{sec:scale} when
there is a continuous spectrum. In Sec.~\ref{sec:disappear} we analyze the
low-energy discrete spectrum when it exists. Classical applications of quantum
mechanics are given in Sec.~\ref{sec:apps}. Finding these applications were, 
in fact, the 
original source of motivation for this work. In Sec.~\ref{sec:other} 
remarks that generalize the inverse-square potential to three dimensions and
$\alpha < -1/4$ are given.

Since this paper is part pedagogical guide, part research, a remark about which
portions are novel is in order. Secs.~\ref{sec:pure}, \ref{sec:break}, and
\ref{sec:renormalize} follow the path set forth in Ref.~\onlinecite{Son}, 
although we provide a different interpretation of the fixed points than in
that work. To the best of our knowledge, Secs.~\ref{sec:prop}, \ref{sec:scale},
\ref{sec:disappear}, and \ref{sec:apps} are new. Lastly, some of the remarks 
made in Sec.~\ref{sec:other} briefly summarize what is already well-established
in the literature regarding the renormalization of the $\alpha < -1/4$ case.

\section{Pure inverse-square potential}
\label{sec:pure}

Does the Hamiltonian given by $H = P^2 + V(X)$ define a sensible quantum
theory? To answer this question one may proceed to construct the physical
Hilbert space of states $\mathcal{H}$ 
over which the operators $X$, $P$, and $H$ are 
self-adjoint. It is important to remember that the self-adjointness property
is not inherent to a differential expression for an operator---one 
must also consider the vector space of functions on which it
acts and the boundary conditions satisfied by those functions. And so it should
be stressed that the physical Hilbert space is not necessarily the space of
square-integrable complex-valued functions $L^2(0,\infty)$ 
equipped with the standard inner product $(\cdot,\cdot)$. 
While this is a bonafide Hilbert space in the functional analysis
sense and is consistent with the degrees of freedom available to a
single spinless particle moving along a line, 
it is not necessarily the physical Hilbert space because certain
functions exist whose behavior at $x = 0$ or $x = \infty$
ruin the hermiticity of observables. Nevertheless, it is certainly true that
$\mathcal{H} \subset L^2(0,\infty)$. We shall construct $\mathcal{H}$ by
forming linear combinations of the eigenfunctions of $H$.
It is crucial that $\mathcal{H}$ be complete with respect to the distance 
function inherited from the inner product so that any Cauchy sequence built 
from elements of $\mathcal{H}$ converge in $L^2$-norm to a limit also in 
$\mathcal{H}$. This is guaranteed, according to Sturm--Liouville theory, 
for a self-adjoint Hamiltonian.\cite{MorsFesh}

The position operator $X$ is already self-adjoint in the space $L^2(0,\infty)$. 
The momentum operator $P$ is self-adjoint if and only if $[f^*g]^\infty_0 = 0$ 
for any $f,g \in \mathcal{H}$. The energy operator $H$ is self-adjoint if
and only if $[f^*g' - {f^*}'g]^\infty_0 = 0$. Both of these conditions arise as
boundary terms resulting from an integration by parts in trying to establish
the equality of $(f,\mathcal{O}g)$ and $(\mathcal{O}f,g)$.

The general solution to the eigenvalue equation
\begin{equation}
\label{eigeq}
  \psi'' = (\alpha/x^2 - E)\psi
\end{equation}
may be constructed as a linear combination of two Frobenius series. For 
$E \neq 0$ and $x \ll |E|^{-1/2}$, the solution is 
$\psi \sim C_+ x^{\nu_+} + C_- x^{\nu_-}$, where
$$
  \nu_\pm = \frac{1}{2} \pm \omega, ~~~~ \omega = \sqrt{\frac{1}{4}+\alpha},
$$
are the roots of the characteristic equation $\nu(\nu - 1) = \alpha$.
We note that the range $-1/4 < \alpha < 0$ implies $0 < \omega < 1/2$, 
so both solutions satisfy $\psi(0) = 0$. We will
see that the regularity of both solutions is what makes the quantum
mechanics of the inverse-square potential so interesting.

It is straightforward to construct eigenfunctions of $H$ using series, and it
turns out that they define ordinary or modified Bessel 
functions. The simplest way to see this is to define a dimensionless variable
$x = |E|^{-1/2}\xi$ and let $\psi = \xi^{1/2}\varphi(\xi)$. 
The resulting ode for $\varphi$ is a variant of Bessel's equation.
For $E < 0$ there is a unique linear combination of the independent solutions 
(with $0 < |C_+/C_-| < \infty$) that
exhibits asymptotic exponential decay and is therefore square-integrable,
namely $\psi_E = x^{1/2}K_\omega(|E|^{1/2}x)$.
Here $K$ is a modified Bessel function of the second kind. 
Such an eigenfunction
is inadmissible as an element of $\mathcal{H}$ for several related reasons.
The scale invariance of the eigenvalue equation implies that if $\psi_E$ is a
normalized state, then so is $\psi_{\lambda^2 E} = \lambda^{1/2}
\psi_E(\lambda x)$ for $\lambda > 0$. 
But that would indicate a continuum of square-integrable states, a situation
which directly contradicts the Feynman--Hellmann theorem as $\lambda$ is not 
an explicit parameter of $H$.\cite{WernCast}
Furthermore, $\psi_E$ and $\psi_{\lambda^2 E}$ are not
orthogonal when $\lambda \neq 1$. Most damning of all, by scaling we can make
$E$ arbitrarily negative so there is no ground state!

The root of these
problems lie in the fact that $\psi_E$ fails to satisfy the required 
boundary conditions. Generally, a solution to Eq.~(\ref{eigeq}) that is
also square-integrable vanishes at spatial infinity and has vanishing
first derivative. In addition, $\psi_E$ also vanishes at the origin.
Therefore, $P$ is self-adjoint with regard to such eigenfunctions. However,
self-adjointness of $H$ requires
\begin{equation}
\label{selfadjoint}
  f^*(0)g'(0) - {f^*}'(0)g(0) = 0.
\end{equation}
But
\begin{align*}
  &
  \lim_{x\downarrow 0}\Bigl[\psi_E(x)\frac{d}{dx}\psi_{\lambda^2 E}(x) 
  - \psi_{\lambda^2 E}(x)\frac{d}{dx}\psi_E(x)\Bigr] \\
  &\propto
  \frac{C_+}{C_-}\omega(\lambda^{\nu_+}-\lambda^{\nu_-}) \neq 0.
\end{align*}
Essentially, hermiticity fails because both near-origin solutions $x^{\nu_\pm}$
are acceptable. One can force the self-adjointness of $H$ by artificially 
introducing another boundary condition to select one solution or the other,
or by modifying the near-origin behavior of the potential.

For $E > 0$ a continuum of eigenfunctions exist, 
$$
  e_k(x) = (kx)^{1/2} J_{\pm\omega}(kx), ~~ k = E^{1/2}.
$$
Both signs satisfy the important closure relation 
$\int_0^\infty e_k(x)e_k(y)dk = \delta(x-y)$ since the 
order of the Bessel function is greater than $-1/2$.\cite{Arfken}
Since both positive
and negative orders give independent representations of the identity operator, 
we have the freedom to
use either sign in forming physical states. That is, consider the subspace
of $L^2(0,\infty)$ whose elements may be expressed as
$$
  f(x) = \int_0^\infty \widetilde{f}(k)e_k(x)dk.
$$
Here we assume that $\|f\| = \|\widetilde{f}\| = 1$ so that the integral
exists by Plancherel's theorem. This describes an inverse Hankel transform.
A function constructed by this kind of superposition inherits the same 
near-origin behavior as that of the Bessel $J$ function. To wit, near the 
origin, $e_k(x) \sim (E^{1/2}x)^{\nu_\pm}$. Notice that the dependence 
on $E$ and $x$ is separable. Therefore, in an integral of the form 
$\int_0^\infty g(E)e_k(x)dE$, where the $x$-dependence is parametric, the 
result will be proportional to $x^{\nu_\pm}$. This
ensures that $H$ is self-adjoint and that one obtains two possible Hilbert 
spaces $\mathcal{H}_\pm$ depending on the
sign chosen in the order of the Bessel function. A final remark: $E = 0$ is
the infimum of the eigenvalues, but strictly speaking
there is no zero-energy eigenstate.

There is a simple physical argument that explains why the Hilbert space is
not unique.\cite{Beane} 
Imagine scattering a de Broglie wave of fixed energy $E > 0$
coming from $x = \infty$. This is somewhat artificial since we should really
speak in terms of a normalizable wave packet, but such extra rigor 
does not change the essential conclusion. 
For $x \gg k^{-1}$ the wavefunction takes the form
$e^{-ikx} + re^{ikx}$. It should be possible to express the reflection
amplitude $r$ in terms of $C_+/C_-$, however there is no condition that
determines the value of $C_+/C_-$ itself! Ordinarily, a boundary condition
like $\psi(0) = 0$ suffices to fix the ratio of coefficients of linearly
independent solutions. In this situation, $\psi(0) = 0$ rules out neither 
$x^{\nu_+}$ nor $x^{\nu_-}$. 

\section{Breaking scale invariance}
\label{sec:break}

Our goal is to connect the two scale-invariant theories
$\mathcal{H}_\pm$ in the Wilsonian sense by linking them in a continuous space
of Hamiltonians. However, the kind of explicit symmetry breaking needed in
$V$ must take place at $x = 0$ and must be sufficient to resolve the
singularity. When framed as a two-body problem in three dimensions, 
it is obvious that ``near-origin'' is synonymous with
``short-distance'' (i.e., nearly coincident particles). 

Before moving on it is worth mentioning what
happens if $V$ is modified at long distances instead of short distances. 
For example, one
analytically attractive method is to add a harmonic trap, $\Omega^2 x^2$. Here
$\Omega^{-1/2}$ serves as an explicit length scale. Miraculously, there exist
raising and lowering operators that create two entirely independent ladders of
states with equally spaced rungs.\cite{WernCast} 
Or, imagine a hard wall at some position
$x = L$. This imposes a quantization of energies related to the zeros of the
Bessel $J$ function. The zeros may be those of $J_{+\omega}$ or 
$J_{-\omega}$, which are, in general, distinct.

\subsection{Regularizing with a square well}

Modify the potential in Eq.~(\ref{V}) so that it reads
\begin{equation}
\label{sqwell}
  V(x) = \biggl\{\begin{array}{ll}
  -g/(bx_0)^2 & 0 < x < bx_0 \\
  \alpha/x^2 & x > bx_0
  \end{array},
\end{equation}
where $g > 0$ and $b > 0$ are dimensionless parameters, and we regard $x_0$ 
as a fixed length scale. 
As usual, hermiticity of the momentum and kinetic energy operators $-id/dx$ and
$-d^2/dx^2$ constrain a wavefunction and its derivative to exist and be 
everywhere continuous, even at the jump discontinuity in $V$.\cite{Branson}

Our expressions are naturally stated in terms of a dimensionless wavenumber,
$$
  \xi = bx_0 |E|^{1/2}.
$$

\subsection{Discrete eigenfunctions}

Let $E < 0$. The exact bound state eigenfunction is
$$
  \psi_{E<0} = \biggl\{\begin{array}{ll}
  A\sin(x\sqrt{E + g/b^2x_0^2}) & 0 \leq x < bx_0 \\
  C\sqrt{|E|^{1/2}x}K_\omega(|E|^{1/2}x) & x > bx_0
  \end{array}.
$$
The allowed energies follow from
\begin{equation}
\label{implicitbd}
  \sqrt{g-\xi^2}\cot\sqrt{g-\xi^2} = \frac{1}{2} 
  + \xi\frac{K_\omega'(\xi)}{K_\omega(\xi)}.
\end{equation}
Constants $A$ and $C$
are fixed by continuity and normalization.
Imagine graphing each side of Eq.~(\ref{implicitbd}) with respect to $\xi$:
the left side is monotone increasing but the right side is monotone decreasing.
Thus, a root occurs only if the left 
side starts somewhere below, or at, the starting point of the right side. 
By making
use of the identity $\lim_{\xi \downarrow 0} \xi K_\omega'(\xi)/K_\omega(\xi) 
= -\omega$, the root $\xi = 0$ obtains when $\sqrt{g}\cot\sqrt{g} = \nu_-$.
Let us denote the solution to this equation by $g_-$. Therefore, a valid
bound state exists when $g > g_-$.

\subsection{Continuum eigenfunctions}

Let $E > 0$. The exact eigenfunction is
$$
  \psi_{E>0} = \Biggl\{\begin{array}{ll}
  A\sin(x\sqrt{E + g/b^2x_0^2}) & 0 \leq x < bx_0 \\
  C_+\sqrt{E^{1/2}x}J_\omega(E^{1/2}x) & \\
  + C_-\sqrt{E^{1/2}x}J_{-\omega}(E^{1/2}x) &  x > bx_0 
  \end{array}. 
$$
The implicit energy equation is
\begin{equation}
\label{implicit}
  \sqrt{g+\xi^2}\cot\sqrt{g+\xi^2} = \frac{1}{2} + 
  \xi\frac{
  C_+J_\omega'(\xi) + C_-J_{-\omega}'(\xi)}{
  C_+J_\omega(\xi) + C_-J_{-\omega}(\xi)},
\end{equation}
where primes denote derivatives with respect to the whole argument.
It is important to note that 
although we have chosen to write these expressions using the notation of
Bessel functions, the analyses in this article rely on little more than
the first couple terms in their power series for small argument.

Suppose we wish either $C_+$ or $C_-$ to vanish. Using the identity 
$\lim_{\xi \downarrow 0}\xi J_\omega'(\xi)/J_\omega(\xi) = \omega$,
Eq.~(\ref{implicit}) becomes, as $b \downarrow 0$, 
$\sqrt{g}\cot\sqrt{g} = \nu_\pm$.
Denote the roots of this equation as $g_\pm$. That is,
\begin{align*}
  C_- = 0: & ~~\sqrt{g_+}\cot\sqrt{g_+} = \nu_+ \\
  C_+ = 0: & ~~\sqrt{g_-}\cot\sqrt{g_-} = \nu_-.
\end{align*}
Since $\sqrt{g}\cot\sqrt{g}$ is monotonically decreasing for
$0 < g < \pi^2$ it follows that $g_+ < g_-$.
For instance, if $\alpha = -3/16$, then $g_+ \approx 0.7136$ and 
$g_- \approx 1.9411$.

The explicit formula for the ratio $C_+/C_-$ is
$$
  \frac{C_+}{C_-}(g,\xi)
  = \frac{-\xi J'_{-\omega}(\xi) + J_{-\omega}(\xi)
  \bigl[\sqrt{g+\xi^2}\cot\sqrt{g+\xi^2}-\half\bigr]}{\xi J'_\omega(\xi) 
  - J_\omega(\xi)\bigl[\sqrt{g+\xi^2}\cot\sqrt{g+\xi^2}-\half\bigr]}.
$$
Assume that $E$ is fixed. For any desired positive value of $C_+/C_-$ there is
a corresponding value of $g \in (g_+, g_-)$. 
We will primarily be interested in the analytical form of
the ratio for small $b$, or equivalently, $\xi \ll 1$. Then 
\begin{equation}
\label{CC}
  \frac{C_+}{C_-} =
  \frac{-2^{2\omega}\Gamma(1+\omega)}{\Gamma(1-\omega)}
  \xi^{-2\omega}
  \frac{\sqrt{g}\cot\sqrt{g} - \nu_-}{\sqrt{g}\cot\sqrt{g} - \nu_+}[1 + o(b)].
\end{equation}
We have established that for $g_+ < g < g_-$ every eigenfunction $\psi_{E>0}$ 
will involve a well-defined linear combination of the fundamental solutions 
characterized by a finite and nonzero value for $C_+/C_-$.
In particular, $C_+/C_-$ scales as $E^{-\omega}$ for fixed $b$ and $g$. 

\section{Renormalizing the coupling}
\label{sec:renormalize}

\subsection{$C_+/C_-$}
\label{sec:CC}

Consider a point in the space of Hamiltonians parametrized by the $(b,g)$-plane.
We would like to preserve the quantity $C_+/C_-$ as $b$ is taken to zero. In 
order to do this, $g$ must vary too. The physical motivation
for preserving $C_+/C_-$ is, as will be discussed later, its direct relation
to the scattering phase shift and indirect relation to the binding energy. These
are low-energy observables that may be measured at spatial infinity.

Take Eq.~(\ref{CC}) for fixed $E$ and small nominal $b$. As $b$ 
decreases further, the factor $\xi^{-2\omega}$ increases, and so $C_+/C_-$
remains constant only if $(\sqrt{g}\cot\sqrt{g}-\nu_-)/(\sqrt{g}\
\cot\sqrt{g}-\nu_+)$ approaches zero from below.

Shrinking $b$ ought to be understood as a flow to the infrared. Why? The 
position of a particle at $x$ can be said to be neither close to the origin
nor far unless a comparison is made to some length scale. $V$ as given by
Eq.~(\ref{V}) lacks such a scale, but introducing the cutoff $bx_0$ in 
Eq.~(\ref{sqwell}) makes it possible to judge whether a physical distance is 
small or large. For instance, for a nominal value of $b = 1$, 
$x/x_0 \ll 1$ indicates that the particle
is ``close'' to the origin, 
while $x/x_0 \gg 1$ indicates that it is ``far.'' In three
dimensions, for a fixed
spatial separation $x$ between two particles, rescaling the cutoff from $x_0$
to $bx_0$ makes the distance between the particles larger in units of
the cutoff. That is, $x/(bx_0)$ increases as $b$ decreases. This is 
precisely the regime one must examine to understand the long-distance behavior 
of the interaction.
 
\begin{figure}[h!]
\centering
\includegraphics[scale=1.0]{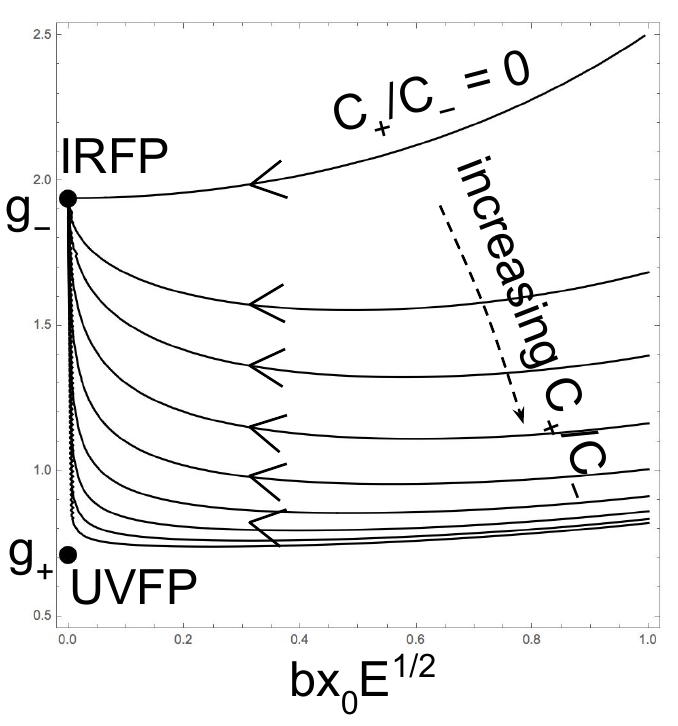}
\caption{Contours of equal $C_+/C_-$ in the $(bx_0E^{1/2},g)$-plane. The
infrared flow ($b \downarrow 0$) is indicated on the contours by arrows. 
The values of $g_\pm$ shown are specific to the example $\alpha = -3/16$, and 
the values of $C_+/C_-$ increase toward the bottom by integer powers of $2$.}
\label{fig:contours}
\end{figure}

The infrared flow described by $b \downarrow 0$ takes a coupling $g$ 
within the interval $(g_+,g_-)$ and makes it tend toward the root of 
$\sqrt{g}\cot\sqrt{g} = \nu_-$. In other words, $g \uparrow g_-$. The precise
manner by which $g$ needs adjustment is called \emph{renormalization}. 
This indicates that $(b,g) = (0,g_-)$ is the \emph{infrared-attractive fixed 
point} (IRFP) of the flow. This is depicted in Fig.~\ref{fig:contours}.

We have learned that the IRFP corresponds to the 
eigenfunction when $C_+ = 0$.\cite{Kaplanconfuse} Why does this make sense? 
Recall that both of the fundamental solutions to the Schr\"odinger equation with
$V$ given by Eq.~(\ref{V}) satisfy the
boundary condition $\psi(0)=0$. Thus, any 
linear combination of the fundamental set is acceptable near the origin.
Once a regulator is introduced (take $b = 1$, say) 
and a unique linear combination is chosen
(up to scalar multiplication), we may ask, in generic terms: How should the 
coefficients of these 
solutions be adjusted so that the appropriate $C^1$-matching can be done 
at the cutoff scale $x_0$? Assuming that $x_0 \ll E^{-1/2}$, the solution
$x^{\nu_-}$ dominates over $x^{\nu_+}$ near the cutoff because 
$\nu_- < \nu_+$. In other words,
$x^{\nu_-}$ changes more rapidly than $x^{\nu_+}$ does for small $x$. As such,
if one desires to have a scattering 
phase shift dictated by the $x^{\nu_+}$ solution,
then one must \emph{finely tune} the coefficient $C_-$ to be zero. Therefore,
from the point of view of $x \gg x_0$, the phase shift is generically 
dictated by the $x^{\nu_-}$ solution, whereas the $x^{\nu_+}$ solution 
is rather special. 

Let us be precise about how $g$ must change as $b$ does. We work in the limit
of small $b$. A certain change of variables makes the analysis 
elegant:\cite{Son}
$$
  \gamma(g) = \sqrt{g}\cot\sqrt{g}.
$$
Then for a given $E$ and $b \ll 1$, 
$$
  \frac{C_+}{C_-} \sim b^{-2\omega}\frac{\gamma - \nu_-}{\gamma - \nu_+}
$$
up to a multiplicative factor independent of $b$.
From the invariance condition $d(C_+/C_-)/db = 0$ we obtain a 
\emph{renormalization group equation}:
\begin{equation}
\label{betafn}
  b\frac{d\gamma}{db} = -(\gamma-\nu_-)(\gamma-\nu_+),
\end{equation}
which is exact in the $b = 0$ limit. Consider $db < 0$. For 
$\nu_- < \gamma < \nu_+$,
the right-hand side of Eq.~(\ref{betafn}), denoted $\beta(\gamma)$ and called
the \emph{beta function}, is positive and so $d\gamma < 0$. 
Thus, $\gamma = \nu_-$ is the IRFP, while $\gamma = \nu_+$ is an 
\emph{infrared-repulsive fixed point}, or, equivalently, an 
\emph{ultraviolet-attractive fixed point} (UVFP). 
This is consistent with what we discovered in terms of the 
parameter $g$. There are two fixed points.\cite{quasi} 

It is both interesting and particularly simple to study the behavior of 
$\gamma$ close to a zero of the beta function. Here a scaling behavior emerges.
Let $\gamma = \gamma(b)$ indicate the coupling associated to some choice of 
$b$. Let $b' = b(1-\epsilon)$ for some infinitesimal $\epsilon > 0$. 
It follows that $b' < b$. Let $\gamma' = \gamma(b')$. 
So $\gamma' \approx \gamma(b) - 
b\epsilon\frac{d\gamma}{db}\bigr|_b
= \gamma - \epsilon\beta(\gamma)$. Differentiate with
respect to $\gamma$ to obtain $d\gamma'/d\gamma \approx 
1 - \beta'(\gamma)\epsilon$.
At a zero $\gamma_*$ of the beta function, the coupling does not change. 
One hypothesizes that the reduced coupling $\gamma - \gamma_*$ obeys a simple
scaling in the vicinity of the zero: $\gamma' - \gamma_* = (\gamma - \gamma_*)
(1-\epsilon)^y$. In the language of the renormalization group, the difference 
$\gamma - \gamma_*$ is called a \emph{scaling variable} and the exponent $y$ 
is called an \emph{RG eigenvalue}.\cite{Cardy} 
But this implies $d\gamma'/d\gamma \approx 1 - y\epsilon$.
Equating both expressions for $d\gamma'/d\gamma$ implies that 
$y = \beta'(\gamma_*)$---the RG eigenvalue is equal to the \emph{slope}
of the beta
function at the fixed point. Specifically, $\beta'(\gamma) = 1-2\gamma$ so
$y = \mp 2\omega$ for $\gamma = \nu_\pm$.

A nice way to summarize what we
have found is to define a \emph{reduced coupling} in the vicinity of each fixed
point. Note that $g_+ < g < g_-$ maps to $\nu_+ > \gamma > \nu_-$.
Near the IRFP define $u = \gamma - \nu_-$. Then
\begin{equation}
\label{IRFP}
  u' = u(1-\epsilon)^{2\omega}.
\end{equation}
Since $u' < u$, we learn that $u$ is an \emph{irrelevant} variable that tends 
to shrink as one enlarges the system. Near the UVFP define 
$u = \nu_+ - \gamma$. Then
\begin{equation}
\label{UVFP}
  u' = u(1-\epsilon)^{-2\omega}.
\end{equation}
Since $u' > u$, we learn that $u$ is a \emph{relevant} variable that tends to
grow.

The assignment of the descriptions ``irrelevant'' and ``relevant'' to 
Eqs.~(\ref{IRFP}) and (\ref{UVFP}) might sound backwards to the reader 
familiar with real-space renormalization, but are, in fact, consistent.
For instance, the real-space approach assigns irrelevant scaling
variables negative RG eigenvalues, not positive ones like ours. 
This is because the sense by which one
progresses to long distances in, say, a discrete lattice model, is a bit 
different. On the lattice there is a spacing $a$ which cannot be adjusted. 
Instead, one coarse-grains over successively larger 
chunks of the (infinite) lattice, each step producing an intermediate lattice 
with larger effective size $a' = (1+\epsilon)a$ for $\epsilon > 0$. Such
procedure reduces the measure of the dimensionless correlation length and is a
way of probing successively longer physical distances. A coupling
obeying the relation $u' = u(1+\epsilon)^{y_\textrm{LAT}}$ for 
$y_\textrm{LAT} < 0$ is then said to be irrelevant because it shrinks. 
This is equivalent to the reduction $b' = (1-\epsilon)b$ used in our analysis
of the inverse-square potential since any physical distance $x$ grows bigger 
in units of the cutoff scale $bx_0$.

Lastly, it is worth reframing the scaling in terms of the original coupling 
$g$. By the same logic as above it should be that, near a fixed point, 
$g' - g_* = (g - g_*)(1-\epsilon)^{\tilde{y}}$ where $\tilde{y} = \beta'(g_*)$.
One should not confuse $\beta(g)$ with $\beta(\gamma(g))$. 
Rather, $\beta(g) = b\frac{dg}{db} = b\frac{dg}{d\gamma}
\frac{d\gamma}{db} = \beta(\gamma)/\gamma'(g)$ by the inverse function theorem
as long as $\gamma'(g) \neq 0$. Differentiating with respect to $g$ yields
$\beta'(g) = \beta'(\gamma) - \beta(\gamma)\gamma''(g)/[\gamma'(g)]^2$. 
However, at a fixed point $\beta(\gamma_*) = 0$ so the extra term vanishes and
we have $\beta'(g_*) = \beta'(\gamma_*)$, or $\tilde{y} = y$. Hence, the 
reduced coupling $g_- - g$ is irrelevant, and $g - g_+$ is relevant with
exactly the same RG eigenvalues as in Eqs.~(\ref{IRFP}) and (\ref{UVFP}).

\subsection{Scattering phase shift}
\label{sec:phase}

One may also frame the renormalization condition as the requirement that 
the relative phase between incoming and outgoing plane waves remain invariant
as one varies the wavelength. Let $\mu = kbx_0$. 
For small $\mu$ we find that $\mu\frac{d\gamma}{d\mu} = \beta(\gamma)$ with 
exactly the same beta function as in Eq.~(\ref{betafn}). 
The salient details are presented in an appendix.

There are two ways to interpret the findings: \textit{(i)} 
One could regard the cutoff $b$ as held fixed at some nominal value and
imagine varying $k$. Observe the phase shift experimentally for some $k$. 
The value so obtained locates a unique point in the $(\mu, g)$-plane.
Taking the long-wavelength limit $k \downarrow 0$ (and hence $\mu \downarrow 0$)
requires adjusting $g$ so that one remains on a certain integral 
curve in this plane.
\textit{(ii)} Another approach is to regard the wavenumber $k$ of the incoming 
plane wave as held fixed, but allow the freedom to adjust $b$.
A choice of $(b,g)$ uniquely specifies a Hamiltonian. For this Hamiltonian, 
there will be a definite phase shift. Now as we take $b \downarrow 0$, 
it is possible to adjust $g$ so that the phase shift does not change if, once
again, we follow an integral curve in the 
$(\mu,g)$-plane. In this interpretation there is a flow between 
Hamiltonians that preserves a long-distance observable.
Thus, taking $b$ to zero is an equivalent way of reaching a long-wavelength 
approximation.

\subsection{Binding energy}

Although our discussion of renormalization has been limited to 
$g \in (g_+, g_-)$, we may also consider $g > g_-$. 
At least one bound state will be present with ground state energy $E$.
One might suspect that $dE/db = 0$ leads to the same renormalization group
equation for $g$, namely Eq.~(\ref{betafn}), as the previously studied
conditions $d(C_+/C_-)/db = 0$ and $dr/db = 0$ in Secs.~\ref{sec:CC}, 
\ref{sec:phase}, and the appendix. 
Our expectation is that this should be true at least in 
the limit $b \downarrow 0$.

On general grounds, the phase shift is 
essentially the phase angle of the reflection amplitude $r$, whereas the 
ground state energy $E$ is the pole of $r$ (more generally, the $S$-matrix) 
in the complex $k$-plane. Since $r$ has modulus one, it is possible to express
it as $r = (s-ik)/(s+ik)$, where $s$ is some real constant and $k$ 
the real wavenumber. However, there is a simple pole at $k = is$, and so
$E = k^2 = -s^2$.
Requiring that $r$ remain constant as $b$ changes implies that the value
of $k/s$ remains constant as $b$ is adjusted. Thus, the renormalization
group equations are identical.

The bound state energy for $g \gtrsim g_-$ is given by solving 
Eq.~(\ref{implicitbd}). For $0 < \omega < 1/2$ and $\xi \ll 1$, 
$\xi K_\omega'(\xi)/K_\omega(\xi) = -\omega 
- 2\frac{\Gamma(1-\omega)}{\Gamma(\omega)}(\xi/2)^{2\omega} + O(\xi^2)$.
Expanding to lowest order in both $\xi$ and $g-g_-$,
\begin{equation}
\label{bindenergy}
  E \sim -C\frac{(g-g_-)^{1/\omega}}{(bx_0)^2},
\end{equation}
where 
$C = [2^{2\omega-2}(1+\alpha/g_-)\Gamma(\omega)/\Gamma(1-\omega)]^{1/\omega}$,
a positive constant that depends only on the value of 
$\alpha$. Eq.~(\ref{bindenergy}) shows once more that $E$ 
remains invariant as long as the reduced coupling $g - g_-$ scales as 
$b^{2\omega}$, the same scaling found when $g$ was just below 
$g_-$. Note that as $b \downarrow 0$, as long as $g-g_-$ follows this scaling
rule, a single bound state (of arbitrary energy) remains!
This is an example of \emph{dimensional transmutation}.

\section{Propagator}
\label{sec:prop}

A choice of $(b,g)$ with $b > 0$ and $g_+ < g < g_-$ selects
a particular Hamiltonian $H_{b,g}$. From this we will now construct the Green's
function---the position-space realization of unitary time evolution. However,
we shall work with pure negative imaginary times, 
$G_{b,g}(x,-it;y) = \ev{x|e^{-tH}|y}$,
$t \geq 0$. This imaginary-time propagator is the solution to 
$\frac{\partial}{\partial t}G = 
\bigl(\frac{\partial^2}{\partial x^2} - V(x)\bigr)G$ with boundary condition
$G(x,-it\downarrow 0;y) = \delta(x-y)$. 
The completeness property of eigenfunctions of
$H_{b,g}$ means that
$$
  G_{b,g}(x,-it;y) = \int_0^\infty dE\, e^{-tE}\psi_E(x)\psi_E(y).
$$
We are motivated to demonstrate the following: two Hamiltonians, one at scale 
$b$ and the other at scale $b'$, but both with couplings close to $g_\pm$, 
will give equivalent long-distance behavior (as measured by $G$ at fixed 
$x,y > bx_0$ and fixed $t$) provided that the coupling $g$ is renormalized from
$b$ to $b'$ according to the scaling laws found in 
Eqs.~(\ref{IRFP}) and (\ref{UVFP}).

At this point we remember that $C_+/C_-$ is not the only ratio needed in order 
to fully specify eigenfunctions. We also need
$$
  \frac{C_-}{A}(g,\xi) = \frac{\sin\sqrt{g+\xi^2}}{\frac{C_+}{C_-}(g,\xi)
  \sqrt{\xi}J_\omega(\xi) + \sqrt{\xi}J_{-\omega}(\xi)}
$$
and the normalization factor $A^2$. This can be fixed by remembering that the
set of eigenfunctions $\{\psi_{E>0}\}$ must satisfy orthonormality and
closure relations. Let us focus on the 
closure property which must hold for any choice of $x,y$ and $g$. Therefore, 
take $x,y < bx_0$ and $g = 0$ so that 
$\int_0^\infty A^2\sin(x\sqrt{E})\sin(y\sqrt{E})dE = \delta(x-y)$. We recognize
here an identity of Fourier sine transforms so it is clear that 
$A^2 = 1/\pi\sqrt{E}$. This is consistent with dimensional analysis since 
$\psi_E$ should have dimensions of $\textrm{length}^{1/2}$. For $g > 0$
we may write
$$
  A^2 = \frac{B(g,\xi)}{\pi\sqrt{E}}
$$
for some dimensionless function $B$ which is strictly positive.
The propagator may be written as
\begin{align}
    & G_{b,g}(x,-it;y) \nonumber \\
  = & \int_0^\infty \frac{BdE}{\pi\sqrt{E}}e^{-tE}
  \Bigl[\frac{C_-}{A}\frac{C_+}{C_-}\sqrt{E^{1/2}x}
  J_\omega(E^{1/2}x) \nonumber \\
  & + \frac{C_-}{A}\sqrt{E^{1/2}x}J_{-\omega}(E^{1/2}x)
  \Bigr]\times[x \leftrightarrow y], ~~~~ x,y > bx_0. \label{propagator}
\end{align}

\begin{figure}[h!]
\centering
\includegraphics[scale=0.8]{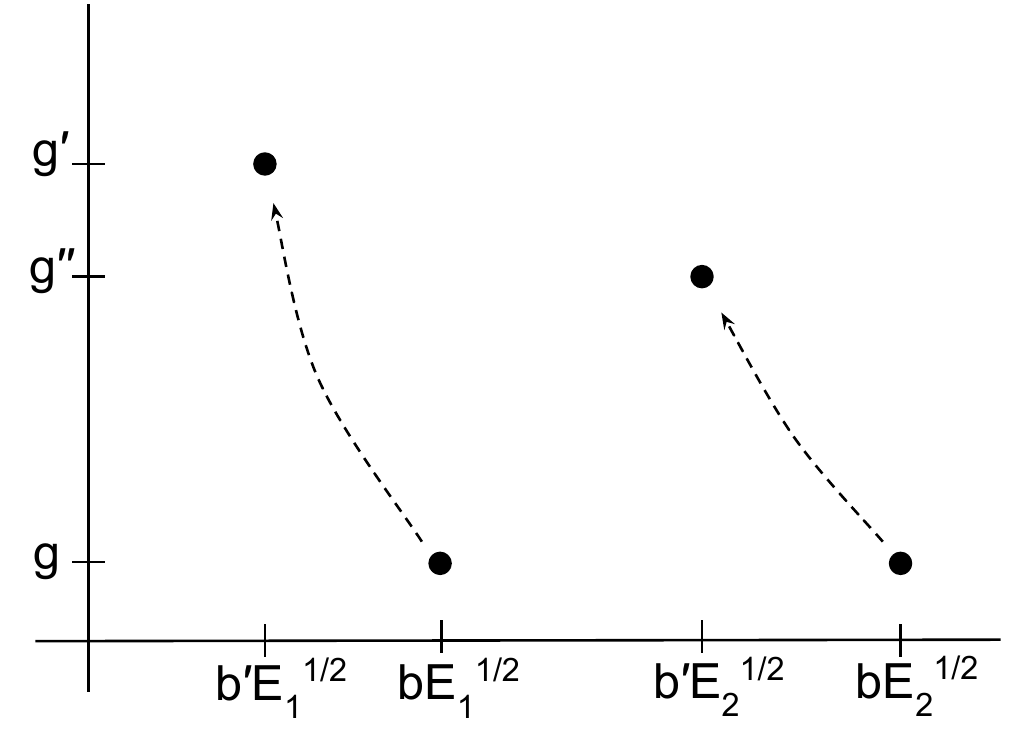}
\caption{A Hamiltonian $H_{b,g}$ has a spectrum corresponding to a horizontal
line in the $(bx_0E^{1/2},g)$-plane. Changing to some $b' < b$ does not
necessarily guarantee that all values of $C_+/C_-$ remain the same for all
energies if we adjust $g$ to a new value. As we illustrate here, 
different values of the coupling, 
$g' \neq g''$, may be needed to keep $C_+/C_-$ unchanged for different energies
$E_1$ and $E_2$.}
\label{fig:vicinity}
\end{figure}

Without explicitly evaluating the integral, we
are interested in proving that $G_{b,g}$ is simply related to $G_{b',g'}$ 
for $b' < b$. The simple relation we seek is an equivalence of the two 
propagators up to an overall scale factor with physical lengths and time 
held fixed. This is a homogeneous transformation with respect to the parameters
$b$ and $g$.
The existence of such a relation would place a severe constraint on the form of
the ratio $C_+/C_-$. The reason is that 
the contours in Fig.~\ref{fig:contours} show exactly how to preserve the 
propagator as $b \downarrow 0$. However, individual
points in Fig.~\ref{fig:contours} are not representative of a choice of 
$(b,g)$. Rather, a given $(b,g)$ corresponds to a horizontal line 
in the plot---all of the uncountably many eigenfunctions of $H_{b,g}$ correspond
to points on this line. Under a shrinking of $b$, one can preserve
the propagator by following the contour passing through each point to its left.
We manage, on an individual basis, to keep all eigenfunctions unchanged 
if we can adjust $g$ so that each eigenfunction's $C_+/C_-$
ratio remains the same. However, there is no guarantee that the new value of
the coupling for one energy will coincide with the new coupling required for
a different energy! More precisely, recall that $C_+/C_-$ is a 
function of two variables $g$ and $\xi$. Write $\xi = b E^{1/2}$ (suppressing
$x_0$) so that $C_+/C_- = f(g,bE^{1/2})$. Consider any
two distinct energies $E_1$ and $E_2$. By changing $b$ to $b'$ it is possible 
to find some $g'$ such that
\begin{equation}
\label{crux1}
  f(g,bE_1^{1/2}) = f(g',b'E_1^{1/2}).
\end{equation}
And it must be possible to find some $g''$ such that 
\begin{equation}
\label{crux2}
  f(g,bE_2^{1/2}) = f(g'',b'E_2^{1/2}).
\end{equation}
These statements are illustrated schematically in Fig.~\ref{fig:vicinity}.
However, $G_{b',g'} \propto G_{b,g}$ requires $g' = g''$, a highly nontrivial 
condition! There is no obvious reason why, starting from generic $g$, the
renormalized couplings $g'$ and $g''$ ought to be the same. Therefore, this 
appears to be an obstruction to finding a simple scaling law for the propagator.

However, it is easy to see that in the vicinity of a fixed point $(0,g_*)$ the
nontrivial condition requiring the equivalence of $g'$ and $g''$ can be 
satisfied. The crux of our argument is this: it is
only close to a fixed point that $f$ takes the asymptotic form
$$
  f(g,bE^{1/2}) \sim \alpha(g-g_*)^\beta(bE^{1/2})^\gamma
$$
for some real numbers $\alpha$, $\beta$, and $\gamma$ which may be extracted
by studying Eq.~(\ref{CC}).
Then Eqs.~(\ref{crux1}) and (\ref{crux2}) become 
$g'-g_* = (b/b')^{\gamma/\beta}(g-g_*)$ and 
$g''-g_* = (b/b')^{\gamma/\beta}(g-g_*)$, respectively. Together they imply
$g' = g''$ as desired.

\subsection{Homogeneous transformation laws}

The scaling relation we seek for the propagator may be obtained from the
following two transformation laws.

\subsubsection{An exact one}

The following equation is exact. Suppose $x,y > bx_0$ and $b' = b/\lambda$ for
some $\lambda > 1$. Then
\begin{equation}
\label{exactlaw}
  G_{b,g}(\lambda x,-i\lambda^2 t;\lambda y) = \lambda^{-1}G_{b',g}(x,-it;y).
\end{equation}

\textit{Proof:} In Eq.~(\ref{propagator}) make a change of variables 
$\tilde{E} = \lambda^2 E$. Within $B$, $C_+/C_-$, and $C_-/A$ this can be 
perfectly compensated by a redefinition of the scale factor $b$ since 
$\xi = bx_0 E^{1/2} = (b/\lambda)x_0\tilde{E}^{1/2} = b'x_0\tilde{E}^{1/2}$.
Since $b' < b$, the appropriate eigenfunctions are still Bessel $J$.

\subsubsection{One valid only close to a fixed point}

The following equation is correct only asymptotically close to a fixed point.
Suppose $x,y > bx_0$, $b \ll 1$, and $\lambda > 1$. Then
\begin{equation}
\label{asymplaw}
  G_{b,u}(\lambda x,-i\lambda^2t;\lambda y) \sim 
  \lambda^{2\nu_\pm - 1} G_{b,u'}(x,-it;y),
\end{equation}
for (upper sign) $u = g-g_+ \ll 1$ and $u' = \lambda^{-2\omega}u$, or
(lower sign) $u = g_- - g \ll 1$ and $u' = \lambda^{2\omega}u$.

\textit{Proof of upper sign case:}
Once again, in Eq.~(\ref{propagator}), make a change of 
variables to $\tilde{E} = \lambda^2 E$. Near a fixed point it is possible 
to absorb factors of $\lambda$ into a redefinition of the reduced coupling 
rather than a redefinition of $b$. Before doing this, replace $C_\pm$ by their 
asymptotic forms for $\xi \ll 1$. (This is justified by taking the 
upper limit of integration to be some large, but finite, value $M/x_0^2$ 
for $M \gg 1$. We shall take $M \to \infty$ later. 
This means that $E^{1/2}$ is bounded above
by $\sqrt{M}/x_0$, and hence, $\xi < b\sqrt{M}$. Let us choose 
$b \ll 1/\sqrt{M}$ so that $\xi$ is small over the entire integration region.)
So in the small-$b$ regime,
\begin{equation}
\label{CA}
  \frac{C_\pm}{A}(g,\xi) \sim 
  \frac{\Gamma(\pm\omega)}{2^{1\mp\omega}}
  \frac{\sin\sqrt{g}(\sqrt{g}\cot\sqrt{g}-\nu_\mp)}{\xi^{\nu_\pm}}.
\end{equation}
Also, we replace $B(g,\xi)$ by its limiting value $B(g_+,0)$ assuming it
exists. Thus,
\begin{align*}
  & G_{b,g}(\lambda x,-i\lambda^2 t;\lambda y) \\ 
  =\, & \frac{1}{\lambda}\int_0^\infty \frac{B(g_+,0)d\tilde{E}}{\pi
  \sqrt{\tilde{E}}} e^{-t\tilde{E}}
  \Bigl[\frac{\mathrm{cst.}\lambda^{\nu_+}}{(bx_0\tilde{E}^{1/2})^{\nu_+}}
  \sqrt{\tilde{E}^{1/2}x}J_\omega(\tilde{E}^{1/2}x) \\
  & + \frac{\mathrm{cst.}u\lambda^{\nu_-}}{(bx_0\tilde{E}^{1/2})^{\nu_-}}
  \sqrt{\tilde{E}^{1/2}x}J_{-\omega}(\tilde{E}^{1/2}x)
  \Bigr]\times[x \leftrightarrow y].
\end{align*}
The constants ``$\mathrm{cst.}$'' appearing above result from expanding 
Expr.~(\ref{CA}) around $g = g_+$. They are nonzero. By writing 
$u = \lambda^{\nu_+-\nu_-}u'$, it is seen that two factors of $\lambda^{\nu_+}$
may be pulled out of the integral. Although this concludes the proof, it is
worth explaining it heuristically. For $g$ close to $g_+$, we can set 
$C_- = 0$ so that the
eigenfunctions are, for small $x$, power laws of the form $x^{\nu_+}$. 
Therefore, in the propagator we have two wavefunctions of the form 
$(\lambda x)^{\nu_+}$ and $(\lambda y)^{\nu_+}$. This is how a factor of 
$\lambda^{2\nu_+}$ is obtained. A proof for the lower sign case is similar.

\section{Scaling}
\label{sec:scale}

By combining Eq.~(\ref{exactlaw}) with (\ref{asymplaw}) we are able to derive
the scaling relation for the propagators evaluated at the same values of 
$x,y,t$, but corresponding to different Hamiltonians $H_{b,g}$ and $H_{b',g'}$.
We shall suppress dependence on the variables $x,y,t$ which are imagined to be
held fixed, and instead highlight the dependence of the propagator on the 
parameters $b$ and $g$. Then
\begin{equation}
\label{propagatorRG}
  G(b,u) \sim \lambda^{-2\nu_\pm}
  G(b/\lambda, u\lambda^{\pm 2\omega}).
\end{equation}
The upper sign is for $g \gtrsim g_+$ and the lower sign for $g \lesssim g_-$.
Eq.~(\ref{propagatorRG}) is a central result of this article. 
One can recognize the RG 
eigenvalues of $u$ found earlier in Eqs.~(\ref{IRFP}) and (\ref{UVFP}).

An equivalent way to state Eq.~(\ref{propagatorRG}) is that the propagator 
satisfies a certain first-order pde which is obtained by 
imagining an infinitesimal change in parameters. 
Let $\lambda = 1 + \delta\lambda$ for 
infinitesimal positive $\delta\lambda$. Expanding to first order in 
$\delta\lambda$, we obtain
\begin{equation}
\label{CallanSymanzik}
  \Bigl[b\frac{\partial}{\partial b} \mp 2\omega u \frac{\partial}{\partial u}
  + 2\nu_\pm\Bigr]G = 0.
\end{equation}
The physical interpretation of Eq.~(\ref{CallanSymanzik}) is that a 
change of $b$ in $G$ can be absorbed through a compensating change in $u$. 
In other words, scale dependency may be exchanged for coupling
dependency, a fact already clear from our derivation of Eq.~(\ref{asymplaw}).
Notice that the coefficients $\mp 2\omega$ of the operator 
$u\frac{\partial}{\partial u}$ are precisely the slope of the beta function 
at the fixed points $\gamma = \nu_\pm$. Eq.~(\ref{CallanSymanzik}) is 
reminiscent of the Callan--Symanzik equation in quantum field theory which
describes how correlation functions change with the renormalization scale.

\subsection{At the fixed points}

We wish to discuss the propagator at the fixed points.
We have seen that $g = g_\pm$ 
corresponds to $C_\mp = 0$ only in the limit that $b$ equals 0. 
Taking this limit squeezes
the sinusoidal part of the eigenfunction into an infinitesimally narrow region 
around the origin. Adapting an integral from Ref.~\onlinecite{PeakInom}, we 
obtain
$$
  G_{0,g_\pm}(x,-it;y) = \frac{\sqrt{xy}}{2t}e^{-(x^2+y^2)/4t}I_{\pm\omega}
  \Bigl(\frac{xy}{2t}\Bigr),
$$
where $I$ is a modified Bessel function of the first kind. Using the 
well-known properties $I_\nu(z) \approx e^z/\sqrt{2\pi z}$ for $|z| \gg 1$, 
and $I_\nu(z) \approx (z/2)^\nu/\Gamma(\nu+1)$ for $|z| \ll 1$, the asymptotic 
behavior is given by 
$$
  G_{0,g_\pm}(x,-it;y) \sim \biggl\{\begin{array}{ll}  
  \frac{1}{\sqrt{4\pi t}}e^{-(x-y)^2/4t} & t \downarrow 0, \\
  \frac{1}{2^{1\pm 2\omega}\Gamma(1\pm\omega)}\frac{1}{\sqrt{t}}
  (\frac{xy}{t})^{\nu_\pm} & t \to \infty
  \end{array}.
$$
The fixed-point theories both reproduce free-particle behavior at short times,
but have differing power-laws at long times.

\subsection{Close to the fixed points}

For $(b,g)$ in the vicinity of $(0,g_\pm)$, but not strictly at those points,
the functional form of the propagator may be uncovered by renormalization
group scaling analysis.\cite{Cardy} We illustrate these standard techniques 
near the UVFP.

\subsubsection{RG for $u > 0$}

For $u = g-g_+ \ll 1$ but strictly greater than zero,
$G(b,u) \sim \lambda^{-2\nu_+}G(b/\lambda, 
\lambda^{2\omega}u)$. Iterating $n$ times gives 
$G(b,u) \sim \lambda^{-2\nu_+ n}G(b/\lambda^n,\lambda^{2\omega n}u)$. 
Since $u$ is relevant, the effective reduced coupling grows under iteration
so $n$ cannot be taken arbitrarily large or else the asymptotic approximation
breaks down. So we stop the iteration at the point 
where $\lambda^{2\omega n}u = u_0$, where $u_0$ is an arbitrary but fixed 
constant that is sufficiently small. Solving yields 
$\lambda^n = (u_0/u)^{1/2\omega}$. Plugging this back in gives 
$G(b,u) \sim (u/u_0)^{\nu_+/\omega}G(b(u/u_0)^{1/2\omega},u_0)$. 
Observe that the parametric dependence of $G$ on $b$ and $u$ has simplified to
the point where we may now express it as
\begin{equation}
\label{propagatorscalinglaw1}
  G(b,u) \sim (u/u_0)^{1 + 1/2\omega} \Phi(b(u/u_0)^{1/2\omega}),
\end{equation}
for some function $\Phi$. At first sight it might appear that the functional
form of $\Phi$ depends on the specific value of $u_0$. Indeed, we could have
imagined the iteration stopping at some $u_0' \neq u_0$. Along the lines of
Eq.~(\ref{propagatorscalinglaw1}) we could then write
$G(b,u) \sim (u/u_0')^{1+1/2\omega}\Psi(b(u/u_0')^{1/2\omega})$ for some 
apparently unrelated function $\Psi$. But the important fact is that the 
left-hand side of Eq.~(\ref{propagatorscalinglaw1}) is insensitive to whether
the scaling variable halts at $u_0$ or $u_0'$. This implies an equality:
$\Psi(z) = (u_0'/u_0)^{1+1/2\omega}\Phi((u_0'/u_0)^{1/2\omega}z)$, and means
that the ratio $u_0'/u_0$, and any powers thereof, 
may be absorbed into a redefinition of the scaling variable $u$. 
Thus, the form of $\Phi$ must be independent of $u_0$. In other words, $\Phi$ is
truly a function with a single argument. In such instance, 
$\Phi$ is referred to as a \emph{scaling function}.

\subsubsection{RG for $u = 0$}

For $u = 0$ our $n$-times-iterated homogeneous transformation law for the 
propagator may be written $G\bigl(\frac{b}{\sqrt{xy}/x_0}, 0\bigr) \sim 
(\lambda^{-n})^{2\nu_+}G\bigl(\frac{b\lambda^{-n}}{\sqrt{xy}/x_0}, 0\bigr)$.
We have reintroduced the spatial variables $x$ and $y$ using dimensional 
analysis. This may be thoroughly justified by redefining the integration 
variable in Eq.~(\ref{propagator}) from $E$ to $E/t$. Without explicitly
evaluating the integral it is easy to see that 
$G = t^{-1/2}f(x/bx_0,x^2/t,y/bx_0,y^2/t)$ for some function $f$. 
Our recursion relation is not sensitive to terms like $x^2/t$ or $y^2/t$; it
must be correct only for $t \gg x^2,y^2$. Furthermore, the propagator must
be symmetric in $x$ and $y$.
Since $\lambda > 1$ one may imagine $n$ being 
so large that $b\lambda^{-n}/(\sqrt{xy}/x_0) = \epsilon$ 
is an exceedingly small fixed 
number. It could be so small that $\epsilon \sqrt{xy}/x_0 \ll 1$
(recall that $x$ and $y$ are fixed). 
This requires that we choose $n \gg \log b/\log \lambda$, which is always
possible for a given $b$ and $\lambda$. So
$$
  G\Bigl(\frac{b}{\sqrt{xy}/x_0},0\Bigr) \sim \epsilon^{2\nu_+}
  \Bigl(\frac{\sqrt{xy}}{bx_0}\Bigr)^{2\nu_+}G(\epsilon,0).
$$
Note that $\epsilon^{2\nu_+}G(\epsilon,0)$ is a constant.
Including an overall $t^{-1/2}$ factor from dimensional
analysis, we obtain
\begin{equation}
\label{propagatorscalinglaw2}
  G\Bigl(\frac{b}{\sqrt{xy}/x_0},0\Bigr) \sim \frac{1}{\sqrt{t}}\Bigl(
  \frac{xy}{b^2x_0^2}\Bigr)^{\nu_+}.
\end{equation}

\subsubsection{Verification}

Fortunately, analytical evaluation of $G$ is rather simple
for asymptotically large $t$. This allows us to check 
Eqs.~(\ref{propagatorscalinglaw1}) and (\ref{propagatorscalinglaw2}).
For large $t$ the integral in Eq.~(\ref{propagator}) is dominated by 
values of $E$ near zero which allows us to replace 
$\sqrt{E^{1/2}x}J_{\pm\omega}(E^{1/2}x)$ by $(E^{1/2}x)^{\nu_\pm}$. 
After rescaling the dummy integration variable to extract $t$, we obtain,
up to an overall constant factor,
\begin{equation}
\label{Ganalytic}
  G_{b,g \sim g_+} \sim \frac{1}{\sqrt{t}}
  \Bigl[\Bigl(\frac{x}{bx_0}\Bigr)^{\nu_+} + c u
  \Bigl(\frac{x}{bx_0}\Bigr)^{\nu_-}\Bigr]
  \times [x \leftrightarrow y],
\end{equation}
where $u = g-g_+ \ll 1$ and $c$ is some constant.

Setting $u = 0$ in Eq.~(\ref{Ganalytic}) gives precisely 
Eq.~(\ref{propagatorscalinglaw2}). It takes a little more effort to show that
Eq.~(\ref{Ganalytic}) is equivalent to Eq.~(\ref{propagatorscalinglaw1}) 
when $u > 0$. Consider
\begin{align*}
  & b^{-\nu_+} + c ub^{-\nu_-} \\
  =\, & (b u^{1/2\omega} u^{-1/2\omega})^{-\nu_+} + c u
  (b u^{1/2\omega} u^{-1/2\omega})^{-\nu_-} \\
  =\, & (bu^{1/2\omega})^{-\nu_+} u^{1/4\omega + 1/2} +
  c(bu^{1/2\omega})^{-\nu_-} u^{1 + 1/4\omega - 1/2} \\
  =\, & u^{1/2 + 1/4\omega}\Phi(bu^{1/2\omega}),
\end{align*}
where $\Phi(z) = z^{-\nu_+} + c z^{-\nu_-}$.
Of course, there is also a similar factor involving $y$. Thus, the factor
$u^{1/2 + 1/4\omega}$ gets squared and becomes $u^{1 + 1/2\omega}$ as desired.

\section{Disappearing bound state}
\label{sec:disappear}

In this section we set $b = 1$ in the regulator Eq.~(\ref{sqwell}) and do not 
allow the well width to vary. For clarity we also take $x_0 = 1$ (it can
always be restored by looking at dimensions). For $g \gtrsim g_-$ the
ground state energy scales as $E \propto (g-g_-)^{1/\omega}$.
As $g \downarrow g_-$, this binding energy vanishes 
at a rate described by the exponent $1/\omega$. 
It must also be the case that the probability to find the particle
within any finite interval tends to zero---the mean position of the particle
(and all higher moments) should run off to infinity in a continuous fashion. 
The rate at which this occurs is given by computing the divergent part of 
$\int x|\psi|^2dx$. In fact, we need only focus on the 
exponentially-falling tail of the wavefunction. We find that
$\int_1^\infty x|\psi|^2dx\bigl/\int_1^\infty |\psi|^2dx \sim
\half|E|^{-1/2}$. Up to a constant factor,
$$
  \ev{x} \sim (g-g_-)^{-1/2\omega}.
$$

The rate of vanishing of $E$, or the degree of divergence of $\ev{x}$, 
is controlled by the long-distance part of $V$. The specific 
scheme chosen for the short-distance part (e.g., square well) has no effect
other than to modify the location of the critical value of $g$. 
This independence of regulator scheme is an example of \emph{universality}: 
physical systems with very different ultraviolet behavior 
(in this case, Hamiltonians with very different excited spectra) 
may have similar low-energy behavior.

A nice exercise is to try a regulator like the linear well,
$V = -gx$ for $0 < x < 1$. Like the square well, it has a single 
dimensionless parameter $g > 0$ responsible for controlling the well slope.
The same scaling exponent $1/\omega$ emerges for the ground state energy as in 
Eq.~(\ref{bindenergy}), but the coefficient $C$ and the location of the fixed 
point $g_-$ are not the same as in the square well case. One says that 
$1/\omega$ is universal, but $C$ and $g_-$ are not.

It is readily proven that the same exponent results for a generic regulator. 
Consider a scheme given by
$$
  V(x) = \biggl\{
  \begin{array}{ll}
  -gf(x) & 0 < x < 1 \\
  \alpha/x^2 & x > 1
  \end{array},
$$
where $f$ is a function defined on $[0,1]$ with additional properties to be
imposed below. 
First we prove that for sufficiently large $g$ the potential binds.
It is sufficient to show that there exists some $\psi \in \mathcal{H}$ with
$(\psi, H\psi) < 0$ since this implies that at least one of the
discrete eigenvalues of $H$ is negative. Note that $\psi$ need not be a state
of definite energy. A convenient choice is $\psi = x\exp(-x/2)$.
So
$$
  \int_0^\infty \psi(-\psi'' + V\psi)dx = \frac{1}{2} 
  - g\int_0^1 f(x) x^2 e^{-x} dx + \frac{\alpha}{e}.  
$$
Assume $f$ is integrable. A negative expectation value is guaranteed for
$g > (\half+\frac{\alpha}{e})/\int_0^1 f(x)x^2 e^{-x}dx$.

Next we prove that for sufficiently small $g$ there is no bound state. 
Recall the fact that the square well regulator, Eq.~(\ref{sqwell}), 
does not produce binding if $g < g_-$, where $g_-$ is entirely determined 
by $\alpha$. So take $g < g_-/f(x)$ for all $x < 1$. We assume $f$ is bounded. 
This means that our potential $V$ is everywhere bounded below by the square 
well. 

The solution to the eigenvalue equation is
$$
  \psi_{E<0} = \biggl\{\begin{array}{ll}
  A\varphi^{(1)}_{g,\epsilon}(x) + B\varphi^{(2)}_{g,\epsilon}(x) & 
  0 \leq x < 1 \\
  C\sqrt{\epsilon^{1/2}x}K_\omega(\epsilon^{1/2}x) & x > 1
  \end{array},
$$
where $\varphi^{(i)}_{g,\epsilon}$ are the linearly independent solutions to 
$\varphi'' + gf(x)\varphi - \epsilon\varphi = 0$, with $\epsilon = -E$.
Suppose that $g$ is large enough to admit at least one bound state so that
$\epsilon$ is a real positive number. Call the critical value of $g$ needed for
this to be true some $g_*$. By continuity,
\begin{equation}
\label{implicitbdgen}
  \frac{{\varphi_{g,\epsilon}^{(1)}}'(1)\varphi_{g,\epsilon}^{(2)}(0)
  -\varphi_{g,\epsilon}^{(1)}(0){\varphi_{g,\epsilon}^{(2)}}'(1)}{
  \varphi_{g,\epsilon}^{(1)}(1)\varphi_{g,\epsilon}^{(2)}(0) 
  -\varphi_{g,\epsilon}^{(1)}(0)\varphi_{g,\epsilon}^{(2)}(1)}
  = \frac{1}{2} + \epsilon^{1/2}\frac{K_\omega'(\epsilon^{1/2})}{
  K_\omega(\epsilon^{1/2})}.
\end{equation}
Call the left side of Eq.~(\ref{implicitbdgen}), $L(g,\epsilon)$. Since
$x = 0,1$ are ordinary points of the ode, 
$\varphi^{(i)}_{g,\epsilon}$ and their derivatives
are finite when evaluated at $x = 0,1$. Now let us consider how $L$ depends
on the parameters. Expanding in energy $\epsilon \ll 1$, we claim that 
$L(g,\epsilon)$ equals $L(g,0)$ plus a term whose order is no larger than 
$\epsilon^{1/2}$. (This is because nonanalyticity in the $\epsilon$-dependence 
of the solutions to the ode develops when $f$ vanishes. For example, if $f = 0$
near $x = 0$, then $\varphi \sim e^{\pm\sqrt{\epsilon}x} \approx 
1 \pm \sqrt{\epsilon}x$.)
Call the right side of Eq.~(\ref{implicitbdgen}), $R(\epsilon)$. 
Expanding in $\epsilon \ll 1$, we obtain
$R(0) + O(\epsilon^\omega)$. Since $\omega < 1/2$, this $O(\epsilon^\omega)$
term dominates over the $O(\epsilon^{1/2})$ term obtained by expanding the
left side. Expanding $L$ once more, but this time in $g$ around $g_*$, we
get $L(g_*,0) + O(g-g_*)$. 
Finally, if $g_*$ is such that $L(g_*,0) = R(0)$, then we have 
$O(g-g_*) = O(\epsilon^\omega)$. It follows that $\epsilon \sim
(g-g_*)^{1/\omega}$. 

\section{Some classical applications}
\label{sec:apps}

\subsection{Brownian motion}

Brownian motion is a stochastic process which may be regarded as the symmetric 
random walk in the limit of infinitesimally small time increments. In a
one-dimensional process, a real random variable $x(t)$ varies with a time 
parameter $t$. The map $x(t)$ is continuous, satisfies an initial condition and
may satisfy a final condition as well. Crucially, the process is such that 
increments in $x$ are independent and identically distributed (i.e., 
the probability distribution for $x(t_2)-x(t_1)$ is independent of $x(t_1)$ for
$t_2 > t_1 > 0$, but they are identical in the sense that 
$\mathbb{P}(x(t_2)-x(t_1) < a) = \mathbb{P}(x(t_1) < a)$). These distributions 
are normal with mean zero and variance equal to twice the time interval. It is 
well known how to rigorously assign a probabilistic measure 
to such a set of real-valued continuous functions. 

\begin{figure}[h!]
\centering
\includegraphics[scale=0.8]{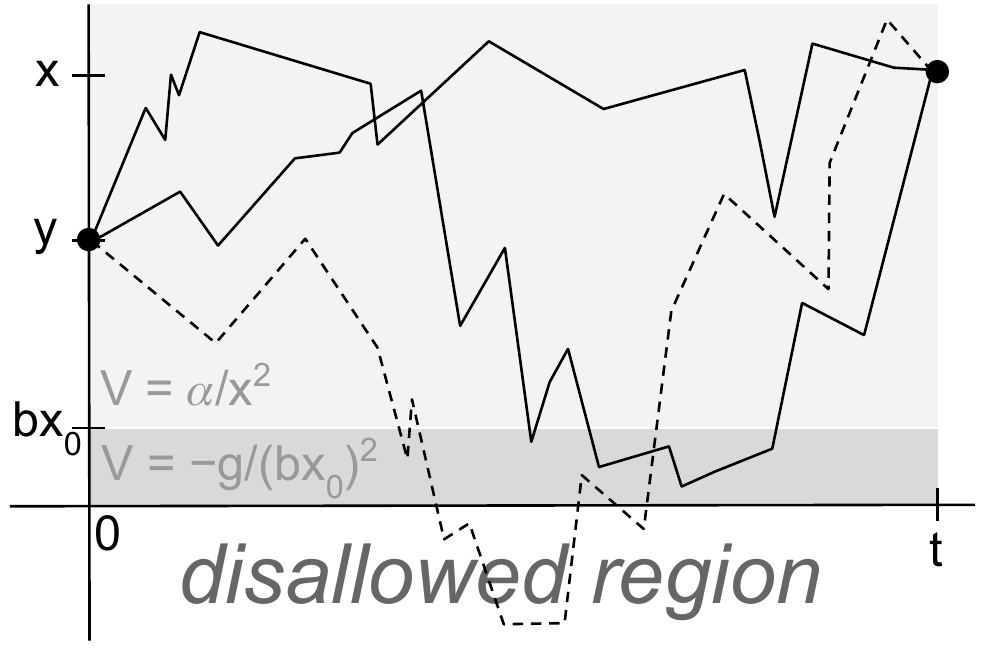}
\caption{The quantity $W$ is given by evaluating the
functional $e^{-\int_0^t V(x(\tau))d\tau}$ over all paths $x(\tau)$ that start
at $y$ at time 0 and end at $x$ by time $t$. Only those that do not ``dip'' into
negative positions are counted. For instance, the dashed path above does not
contribute to the expectation.}
\label{fig:Wiener}
\end{figure}

We are interested in the following conditional expectation 
value,\cite{Wienernorm}
\begin{align*}
  W(x,t;y) & = \mathbb{E}'_{x,t;y,0}\Bigl[e^{-\int_0^t V(x(\tau))d\tau}\Bigr] \\
  & = \lim_{N\to\infty}\int_{-\infty}^\infty dx_1\dotsb
  dx_{N-1}\, (4\pi\epsilon)^{-N/2} \\
  & ~~~~\times\exp\biggl(-\sum_{j=0}^{N-1} 
  \frac{(x_{j+1}-x_j)^2}{4\epsilon} + \epsilon V(x_j)\biggr),
\end{align*}
with $\epsilon = t/N$, $x_0 = y$, and $x_N = x$. Since $V$ is infinite for all 
$x \leq 0$, any path that ``dips'' into $x \leq 0$
at any time will be weighted by $e^{-\infty} = 0$. This effectively eliminates
from consideration all paths that travel left of the origin. Thus, we are 
considering Brownian motion on the half-line $x > 0$. See Fig.~\ref{fig:Wiener}.

\noindent
Kac proved that $W$ satisfies the pde 
$\frac{\partial}{\partial t}W = \bigl(\frac{\partial^2}{\partial x^2} - V
\bigr)W$ with boundary
condition $\lim_{t\downarrow 0}W(x,t;y) = \delta(x-y)$. This is 
mathematically identical to the problem of finding the quantum mechanical 
Green's function in imaginary time.\cite{Schulman} 

Consider $V$ given by Eq.~(\ref{sqwell}) and let $g = g_\pm$.
According to the quantum result Eq.~(\ref{propagatorscalinglaw2}), for 
$x,y > bx_0$, $b \ll 1$, and asymptotically large $t$,
$W(x,t;y) \sim t^{-1/2}(xy/b^2x_0^2)^{\nu_\pm}$. Consider two
Brownian expectations: one for all continuous paths from $y$ to $x$, and 
another from $\lambda' y$ to $\lambda x$, where all positions are strictly 
greater than $bx_0$. Then in the large-$t$ limit,
$$
  W(\lambda x,t;\lambda 'y) \sim (\lambda\lambda')^{\nu_\pm}W(x,t;y).
$$
In particular, observe that $\lambda' = 1/\lambda$ leads to asymptotic
equivalence.
This is rather suprising from the stochastic point of view because the 
interval where the potential depends on $g$ can be made 
arbitrarily narrow, neverthless it exerts an outsized influence on the 
sum over paths! More precisely, consider the set of continuous functions 
that begin at $(y,0)$ and end at $(x,t)$ while always satisfying $x(\tau) > 0$ 
for all $\tau \in (0,t)$. Sampling only from this space, what is the conditional
probability that a particle passes through the portal $0 < x < bx_0$ at some
intermediate time $t_1$? There is a finite answer to this 
question\cite{condBrown} and it can be made arbitrarily 
small by taking $b \to 0$. A naive conclusion might be then that the value of 
$g$ has little effect on $W$. However, the analysis shows the
opposite: $W$ exhibits very different scaling at two special values of $g$.

\subsection{Statistical mechanics of a chain}

Yet another way to interpret the imaginary-time Green's function 
is as a classical partition function.
Consider a discrete and finite one-dimensional system with $N$ real degrees 
of freedom $x_j$. Impose fixed
boundary conditions $x_0 = y$ and $x_{N+1} = x$. 
On the set of $\{x_j\}$ define a probability measure by 
the Boltzmann factor, $\mu = \exp(-S)/Z$, where the energy of a configuration is
$$
  S = \sum_{j=0}^N\Bigl[\frac{1}{4\epsilon}(x_{j+1}-x_j)^2
  + \frac{\epsilon}{2}V(x_j) + \frac{\epsilon}{2}V(x_{j+1})\Bigr].
$$
$\epsilon$ is a nearest-neighbor coupling and $V$ is an external potential. 
We take $\epsilon > 0$ so that the site-to-site coupling is attractive---it
is energetically favorable for $x_j$ to be similar in value to $x_{j\pm 1}$.
Lastly, $Z$ is a constant chosen to satisfy the normalization condition. 
Explicitly,
$$
  Z_{xy}(N,\epsilon) = \int dx_1\dotsb dx_N\, e^{-S}.
$$
A priori there is no relation between $N$ and $\epsilon$.
One is typically interested in evaluating expectations in the measure $\mu$
in the infinite volume limit.

It is well known that, in the double scaling limit
$N \to \infty$, $\epsilon \to 0$ such that $N\epsilon = t$ fixed, there is a
correspondence between the classical statistical mechanics problem we have just
defined and quantum mechanics of a particle propagating in negative imaginary
time.\cite{qmsm} In fact,
$$
  W(x,t;y) \propto Z_{xy}(t),
$$
where the proportionality constant is independent of $g$.
The limit we have taken is a continuum limit.

Now we wish to take a
thermodynamic limit by making the ``volume'' $t$ arbitrarily large.
In this limit we are interested in extracting 
the free energy density defined by 
$f_{xy} = \lim_{t\to \infty}\frac{-1}{t}\log Z_{xy}(t)$. 
According to the discussion in Sec.~\ref{sec:disappear}, for $g > g_*$
a bound state exists so the dominant behavior of $W(x,t;y)$ is 
exponential in $t$ for large $t$. That is, $W(x,t;y) \sim e^{-E_0t}\psi_0(x)
\psi_0(y)$ plus exponentially suppressed corrections. Here $E_0$ is the
ground state energy. Therefore,
$f_{xy} \sim E_0$. Terms like $(\log \psi_0(x))/t$ vanish in the 
large-$t$ limit so the specific choice of $x$ and $y$ do not affect 
the free energy density in the thermodynamic limit. Thus,
$$
  f_{xy} \propto (g-g_*)^{1/\omega}, ~~ g \gtrsim g_*.
$$
This is finite and so an extensive phase exists if $g > g_*$. However, 
if $g < g_*$, then $W(x,t;y)$ is asymptotically a power law in $t$ so the 
system is no longer extensive. In other words,
$$
  f_{xy} \sim 0, ~~ g < g_*.
$$
We see that the loss of extensivity as the parameter $g \downarrow g_*$
is characterized by a universal critical exponent $1/\omega$. 

The phase transition described by the crossing of $g$ across $g_*$ is 
different from the kind of finite-temperature phase transition
familiar from lattice models. One prominent difference is that 
in short-range lattice spin systems, the
free energy is extensive on both sides of the critical point.
Nevertheless, similar extensive-to-nonextensive phase transitions are described
in Ref.~\onlinecite{Nisoli} for the case $\alpha < -1/4$. These authors treat
$\alpha$ as a variable thermal parameter. In one
experimentally realizable system it is explained how the
approach $\alpha \uparrow -1/4$ reproduces Berezinskii--Kosterlitz--Thouless
scaling related to a topological transition between winding states of a 
floppy polymer circling a defect.

\section{Other aspects of the inverse-square potential}
\label{sec:other}

While there are other features and applications of the inverse-square potential,
here we make two brief remarks that extend our analysis and bridge our work
with other pedagogical discussions of the inverse-square 
potential.\cite{Essin, CoonHolstein}

\subsection{Three dimensions}

Our results apply to two particles interacting in the $s$-wave via a 
central potential that is $\alpha/r^2$ at large $r$.\cite{CoonHolstein}
Since the potential is time-independent 
the nontrivial part of the two-particle wavefunction is obtained
by solving $-\frac{\hbar^2}{2m}\nabla^2_r \psi + V\psi = E\psi$, where $m$
is the reduced mass. In a zero-orbital-angular-momentum state, $\psi \propto 
U(r)/r$. It is well known that the radial function $U$ obeys an
equation that looks just like the Schr\"odinger equation on the half-line,
$-\frac{\hbar^2}{2m}d^2U/dr^2 + V(r)U = EU$, $r \geq 0$.

It turns out that the appropriate boundary condition is that $U(0) = 0$,
precisely analogous to that of our one-dimensional example. In 
Ref.~\onlinecite{Shankar} a proof is given by 
demanding that $-d^2/dr^2$ be hermitian with respect to the 
space obtained from all square-integrable linear combinations of functions 
$U$ satisfying the eigenvalue equation. This argument is repeated below.

Let $U_1$ and $U_2$ be two such functions. The combination of square 
integrability and satisfaction of the ode implies that $U_i$ and $U'_i$ 
limit to 0 as $r$ goes to $\infty$. Then operator hermiticity necessitates that 
$(U_1'(0)/U_1(0))^* = U_2'(0)/U_2(0)$, which is precisely
Eq.~(\ref{selfadjoint}) again. From this we see that $U_1(0)$ and $U_2(0)$ 
are equal up to a multiplicative constant for which the only self-consistent 
choice is 1. Thus, all $U_i(0)$ must equal the same constant $c$. However, 
if $c \neq 0$, then $\psi \sim 1/r$ which is problematic because 
$-\nabla_r^2(1/r) \propto \delta^3(\mathbf{r})$ yet
there is no corresponding delta function in $V$. The only way to avoid this
situation is to have $c = 0$.

\subsection{$\alpha < -1/4$}

While the case $\alpha < -1/4$ has been well-studied in the literature (see,
for example, Refs.~\onlinecite{Essin} and \onlinecite{Beane}) 
it is worth highlighting its 
renormalization group analysis for its qualitatively different 
long-distance behavior.

The pure $\alpha/x^2$ potential with $\alpha < -1/4$ has a spectrum unbounded
from below so one really does need to regulate the Hamiltonian with a cutoff
in order to obtain a healthy theory. Suppose this is done with the 
finite square well, Eq.~(\ref{sqwell}). 
Let $\epsilon = -Ex_0^2$. Then for the shallowest
bound states satisfying $\epsilon \ll 1$, the implicit energy equation is
\begin{equation}
\label{limitcycle}
  \sqrt{g}\cot\sqrt{g}\sim \half - |\omega|\tan(|\omega|\log b +
  \half|\omega|\log\epsilon + \phi),
\end{equation}
where the phase $\phi$ is determined by the normalizable solution and
is not a free parameter. This regulator breaks the continuous scale symmetry
to a discrete subgroup (notice that $\epsilon \to e^{-2\pi/|\omega|}\epsilon$
leaves Eq.~(\ref{limitcycle}) invariant). Fix $\epsilon$ and 
consider the locus of points in the 
$(\log b, g)$-plane obeying Eq.~(\ref{limitcycle}). 
If $(\log b_0,g_0)$ is any such
point, then a continuous curve of points exist in its vicinity. By 
following the curve to more negative $\log b$, $g$ tends toward zero.
Eventually, there will come a time when $g$ attains zero, but at this point we
may jump to another curve and start the flow all over again with a finite
value of $g$. This is possible because there are an infinite number of solutions
$g$ for a given $b$. The periodicity of tangent means that the coupling $g$
is renormalized, but never approaches a limit, as $b$ decreases. Instead,
each time $\log b$ is shifted by $-\pi/|\omega|$, $g$ may be identified with
its starting value. Such identification is natural because it keeps $g$
positive. The process may be replicated indefinitely; 
$b$ remains strictly positive because it is reduced by a multiplicative
factor. This describes a renormalization group \emph{limit cycle}---the coupling
$g$ cycles through an interval of values.

\bigskip

\noindent
\textbf{Note added:}
After publication we became aware of a previous RG study in which all possible
universality classes are exhibited for the line-depinning phase transition in
arbitrary dimensions for an attractive pinning potential with an inverse-square
tail.\cite{Kolo} In fact, Ref.~\onlinecite{Kolo} predates Ref.~\onlinecite{Son}.
The scope of Ref.~\onlinecite{Kolo} is far more broad than the simple question
taken up in our Secs.~\ref{sec:disappear} and \ref{sec:apps}B, although the
criterion used to signal a phase transition---the disappearance of a bound 
state---is precisely the same. Our conclusion in Sec.~\ref{sec:apps}B may be
found in Sec.~IVB of Ref.~\onlinecite{Kolo}. Among the many detailed discussions
in that work is a very natural example of a classical statistical mechanical
application of the one-particle quantum problem, namely the attraction of a 
(one-dimensional) wetting liquid interface to an impenetrable substrate. We
thank Eugene Kolomeisky for bringing this to our attention.

\appendix*

\section{Integral curves of constant phase shift}

For $k = E^{1/2} > 0$, the time-independent part of the 
solution to the Schr\"odinger equation with $V$
given by Eq.~(\ref{sqwell}) may be written
\begin{align*}
  \psi &= \sqrt{\frac{\pi}{2}}e^{-i\omega\pi/2-i\pi/4} \\
  &~~~\times
  \left\{\begin{array}{ll}
  A\sin(x\sqrt{k^2 + g/b^2x_0^2}) & 0 \leq x < bx_0 \\
  \sqrt{kx} H^{(2)}_\omega(kx) 
  + r e^{i\omega\pi+i\pi/2}\sqrt{kx} H^{(1)}_{\omega}(kx)
  & x > bx_0
  \end{array}\right..
\end{align*}
$H^{(i)}_\omega$ are Hankel functions.
The asymptotic form of this solution is $\psi \sim e^{-ikx} + r e^{ikx}$.
Doing the matching at $x = bx_0$ yields the following explicit formula
for the reflection amplitude,
$$
  r = ie^{-i\omega\pi}\frac{1 - ic}{1 + ic}, ~~~~
$$
where
\begin{align*}
  c &= \frac{Y'_\omega(\mu)-dY_\omega(\mu)}{J'_\omega(\mu)-dJ_\omega(\mu)}, \\ 
  d &= \frac{2\sqrt{\mu^2 + g}\cot\sqrt{\mu^2 + g} - 1}{2\mu}.
\end{align*}
It is obvious that $|r| = 1$ so this is pure phase.

Demand that $dr/d\mu = 0$. This is only possible if $g$ is allowed 
to ``run'' with $\mu$. Therefore, the $\mu$ dependence of the phase $r$ is
present both explicitly and via implicit dependence through $g$. 
Write $r = R(\mu, g(\mu))$. Then
$$
  0 = \frac{dr}{d\mu} = \frac{\partial R}{\partial \mu} + 
  \frac{\partial R}{\partial g}\frac{\partial g}{\partial \mu}.
$$
This is a linear first-order pde of the form 
$R_\mu + f(\mu,g)R_g = 0$, with 
$$
  f = \frac{-\alpha\zeta\sin^2\zeta + g\zeta\cos^2\zeta - g\sin\zeta\cos\zeta}{
  (\mu/2)(\zeta - \sin\zeta\cos\zeta)}, ~~~~
  \zeta = \sqrt{\mu^2+g}.
$$
The directional derivative of $R$ in the direction of the vector field 
$(1,f(\mu,g))$ is zero. Therefore, in the $(\mu,g)$-plane there is a 
one-parameter family of solutions $\phi(\mu,g) = c$ for constant $c$.
Since $R$ is constant along each integral curve, this means
that $R$ depends only on the value of $c$, or, equivalently, all the $\mu$ and
$g$ dependence in $R$ appears as $r = R(c) = R(\phi(\mu,g))$.

Define the phase shift $\delta$ by $r = -e^{2i\delta}$.
The rationale for this definition is that for scattering off an attractive
potential, $\delta$ will be positive and 
the wave function is ``drawn into'' the well by a distance given by
$\delta/k$ relative to the case of scattering off a hard wall.
When the de Broglie wavelength of the incident particle is much longer than
the width of the square well, $k^{-1} \gg bx_0$ (i.e., $\mu \ll 1$),
$$
  \delta = \frac{\pi}{4}(1-2\omega) - 
  \frac{\pi}{\omega(2^\omega\Gamma(\omega))^2}
  \frac{\sqrt{g}\cot\sqrt{g} - \nu_+}{\sqrt{g}\cot\sqrt{g} - \nu_-}
  \mu^{2\omega} + O(\mu^{4\omega}).
$$

\begin{acknowledgments}

We are indebted to Laurence Yaffe for a careful reading of the manuscript and
for helping to clarify the use of RG arguments that lead to scaling functions.
We gratefully acknowledge helpful correspondence with David Kaplan, 
Cristiano Nisoli, and Lawrence Schulman.

\end{acknowledgments}


\begin{thebibliography}{99}
\bibitem{Kaplanconfuse}
Strangely, in Ref.~\onlinecite{Son} it is claimed that the $C_- = 0$ solution
should be identified with the IRFP. See p.~6 in the paragraph below Eq.~(13) and
also p.~8 in the paragraph below Eq.~(19).
\bibitem{quasi} 
Note that $\gamma(g)$ is monotonically decreasing for $0 < g < \pi^2$.
However, it is quasiperiodic in the sense that the shape of its graph repeats
for intervals $(\pi^2,(2\pi)^2)$, $((2\pi)^2,(3\pi)^2)$, etc. Since 
$0 < \nu_\pm < 1$, there are a pair of fixed points in each of these intervals.
\bibitem{Wienernorm} 
It should be noted that $W$ is not quite a true expectation value
because it is not properly normalized. The correctly normalized expression
turns out to be $W(x,t;y)/\rho(x,t)$, where $\rho(x,t) = e^{-x^2/4t}/
\sqrt{4\pi t}$. See Ref.~\onlinecite{Schulman}.
\bibitem{condBrown} For some $0 < t_1 < t$, the probability is given by 
$\int_0^{bx_0}\rho(x_1,t_1)\rho(x-x_1,t-t_1)dx_1/
\int_0^\infty\rho(x_1,t_1)\rho(x-x_1,t-t_1)dx_1$, where $\rho$ is given in
Ref.~\onlinecite{Wienernorm}.
\bibitem{qmsm}
Here is a reminder of how that correspondence works. Consider
the operator mechanics given by the hermitian transfer matrix
$T = \exp[-\epsilon V(X)/2]\exp[-\epsilon P^2]\exp[-\epsilon V(X)/2]$, where
$[P,X] = -i$. Then $\ev{x|T^{N+1}|y}$ equals, using the completeness property 
for $P$ and $X$, $(2\pi\epsilon)^{-(N+1)/2}\int dx_1 \dotsb dx_N\, e^{-S}$. 
But $\lim_{\epsilon \to 0}\frac{-1}{\epsilon}\log T = P^2 + V(X)$. Thus,
$\ev{x|e^{-t(P^2+V(X))}|y} = (2\pi\epsilon)^{-(N+1)/2}Z_{xy}(t)$.
We recognize the left-hand side as $G(x,-it;y)$. A trace over positions is not
taken.
\end{thebibliography}
\end{document}